\documentclass{PoS}

\usepackage{amsmath}
\usepackage{color}
\usepackage{subfig}

\newcommand{\ncteq}{{\tt nCTEQ\!\! }}

\vbox{
\null \vspace{0.3in}
\hbox{SMU-HEP-14-03}
}

\title{Update on nCTEQ PDFs: nuclear PDF uncertainties and LHC applications}

\ShortTitle{nCTEQ nuclear PDFs}

\author{\speaker{A. Kusina}$^a$, K.~Kova\v{r}\'{\i}k$^b$, T.~Je\v{z}o$^c$,
        D.~B.~Clark$^a$, F.~I.~Olness$^a$, I.~Schienbein$^d$, J.~Y.~Yu$^a$\\
        \llap{$^a$}Southern Methodist University, Dallas, TX 75275, USA\\
        \llap{$^b$}Institut f{\"u}r Theoretische Physik, Westf{\"a}lische Wilhelms-Universit{\"a}t M{\"u}nster,
                   Wilhelm-Klemm-Stra{\ss}e 9, D-48149 M{\"u}nster, Germany\\
        \llap{$^c$}Department of Physics, University of Durham, Durham DH1 3LE, UK\\
                   Department of Mathematical Sciences, University of Liverpool, Liverpool L69 3BX, UK\\
        \llap{$^d$}Laboratoire de Physique Subatomique et de Cosmologie,
                   Universit\'e Grenoble-Alpes, CNRS/IN2P3,
                   53 avenue des Martyrs, 38026 Grenoble, France\\
        E-mail: \email{akusina@smu.edu}, \email{karol.kovarik@uni-muenster.de},
                \email{T.Jezo@liverpool.ac.uk}, \email{dbclark@smu.edu}, \email{olness@smu.edu},
                \email{ingo.schienbein@lpsc.in2p3.fr}, \email{yu@physics.smu.edu}}

\abstract{We present updated nCTEQ nuclear parton distribution functions with errors
          including pion production data from RHIC. We compare them with the results
          of other groups and present selected LHC applications.}

\FullConference{XXII. International Workshop on Deep-Inelastic Scattering and Related Subjects,\\
		28 April - 2 May 2014\\
		Warsaw, Poland}

\begin{document}

\section{Introduction}
\label{sec:intro}

Nuclear parton distribution functions (PDFs) are an important ingredient
required for the description of heavy ion collisions done at the LHC and RHIC.
They are also important for the precise determination of the free-proton PDFs,
as some of the data used in the free-proton fits are obtained using
nuclear targets, e.g. neutrino DIS experiments.
In this contribution we report on the updated analysis of the \ncteq nuclear
parton distributions and mention some of its application to LHC physics.

In the global analysis presented below we are using the same framework
(parametrization, etc.) as in the previous \ncteq
fits~\cite{Schienbein:2007fs,Schienbein:2009kk,Kovarik:2010uv,Stavreva:2010mw}
with the addition of an error analysis. The free-proton baseline PDFs~\cite{Owens:2007kp}
are based on the CTEQ6.1 proton fit~\cite{Stump:2003yu}.
For the error estimate we use the Hessian method that was introduced in
refs.~\cite{Pumplin:2000vx,Pumplin:2001ct}.
This framework was already briefly discussed in the DIS2013 proceedings~\cite{Kovarik:2013sya}
and we refer the reader to that text and to refs.~\cite{Schienbein:2009kk,Kovarik:2010uv}
for more details about it.

The new ingredient presented in the current paper is the addition of the single pion
production data from RHIC~\cite{Adler:2006wg,Abelev:2009hx}.
The results shown below are still preliminary; the official release of the \ncteq
PDFs will be available later this year.

\section{Global analysis}
\label{sec:analysis}
The bulk of the data used in the current analysis is DIS and DY data
on different nuclear targets,%
    \footnote{For details on which data sets are used see Tables I,
    II and III in ref.~\cite{Schienbein:2009kk}. Note also that we
    do not include here any of the neutrino DIS data.}
which after kinematical cuts gives 708 data points. Additionally we
include 32 data points of $\pi^0$ production at
RHIC~\cite{Adler:2006wg,Abelev:2009hx}.%
    \footnote{We used the following kinematical cuts: for DIS and DY data
    $Q>2$ GeV and $W>3.5$ GeV, and for the single pion production
    $p_T\in(1.7,15.5)$ GeV.}
All of these data are included with weight of one in the global $\chi^2$ function.

\begin{figure*}[t]
\centering{}
\includegraphics[clip,width=0.48\textwidth]{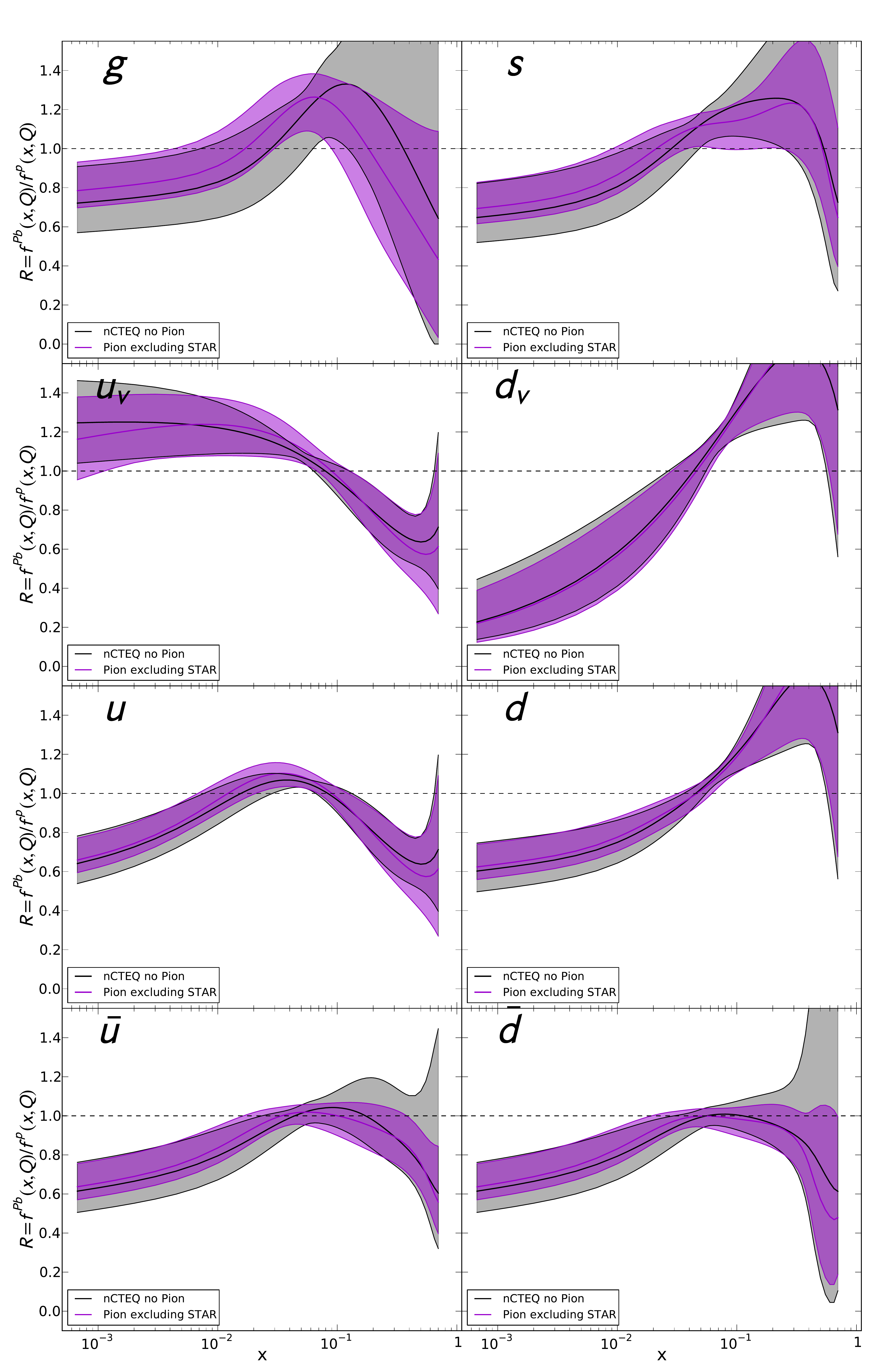}
\quad{}
\includegraphics[width=0.48\textwidth]{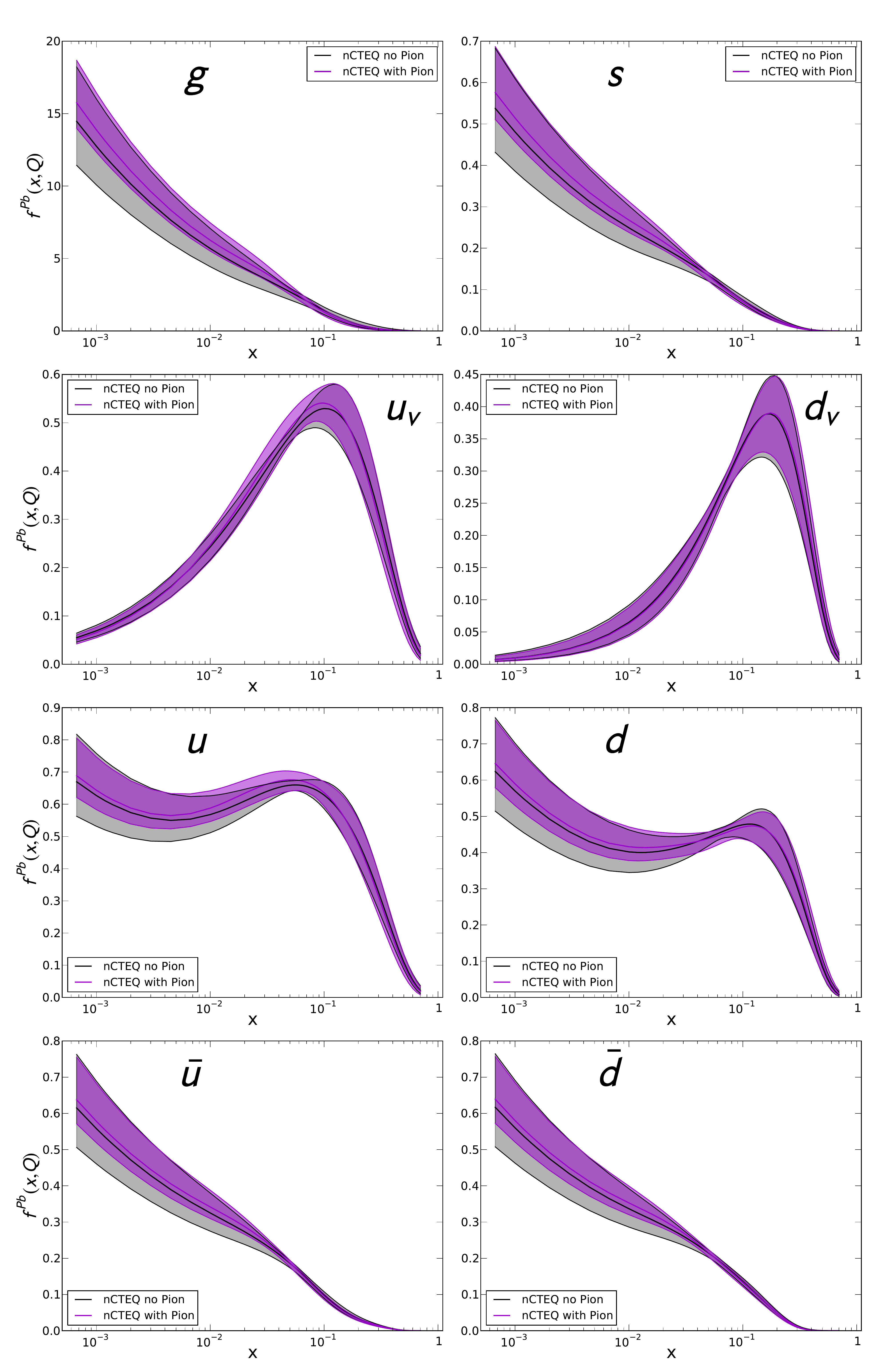}
\caption{Comparison of \ncteq fits with (violet) and without (gray) pion data.
On the left we show nuclear modification factors defined as ratios of proton PDFs
bound in lead to the corresponding free proton PDFs, and on the right we show the
actual bound proton PDFs for lead. In both cases scale is equal to $Q=10$ GeV.
}
\label{fig:compar_pion-vs-NOpion}
\end{figure*}

We present here two fits: (i) excluding the pion data, and (ii) including
the pion data. For both of them we used 16 free fitting parameters, and in both
cases it allowed us to obtain very good fits with $\chi^2/dof$ respectively
equal to 0.87 (without pion) and 0.85 (with pion).
In figure~\ref{fig:compar_pion-vs-NOpion} we compare the two fits.
We show there nuclear correction factors (left panel) and PDFs themselves
(right panel) in case of lead at a scale of $Q=10$ GeV.
We can see that, as expected, the biggest impact of the pion data is
seen in case of the gluon PDF where there is a shape change.
Additionally, if we compare gluon PDF of the above fits with the gluon from the
EPS09 fit~\cite{Eskola:2009uj} (where pion data~\cite{Adler:2006wg} was
included with weight of 20) we will see that the gluon of the fit with pion data
is closer to the EPS09 one.
%
%

\begin{figure*}[t]
\centering{}
\includegraphics[clip,width=0.48\textwidth]{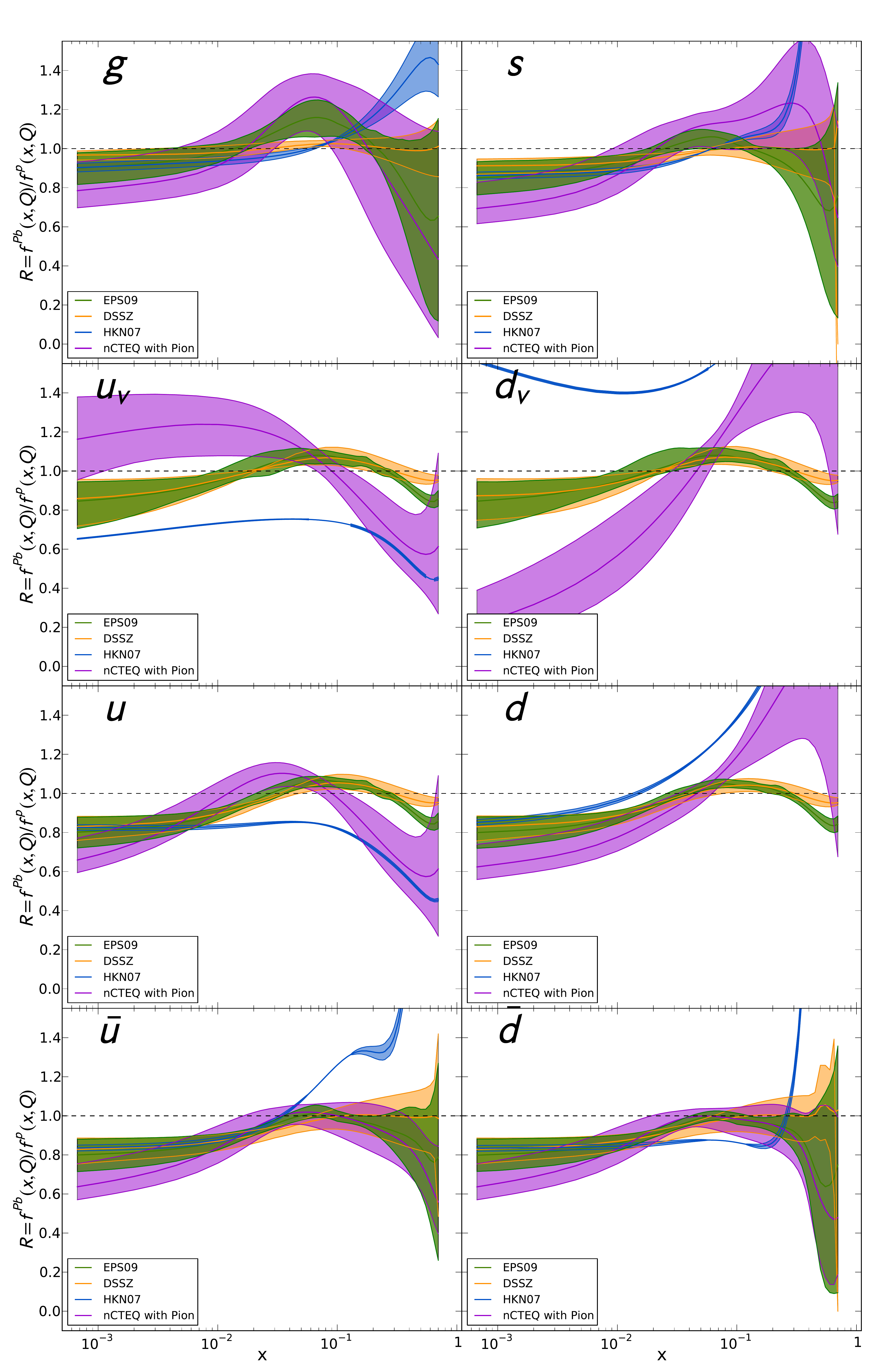}
\quad{}
\includegraphics[width=0.48\textwidth]{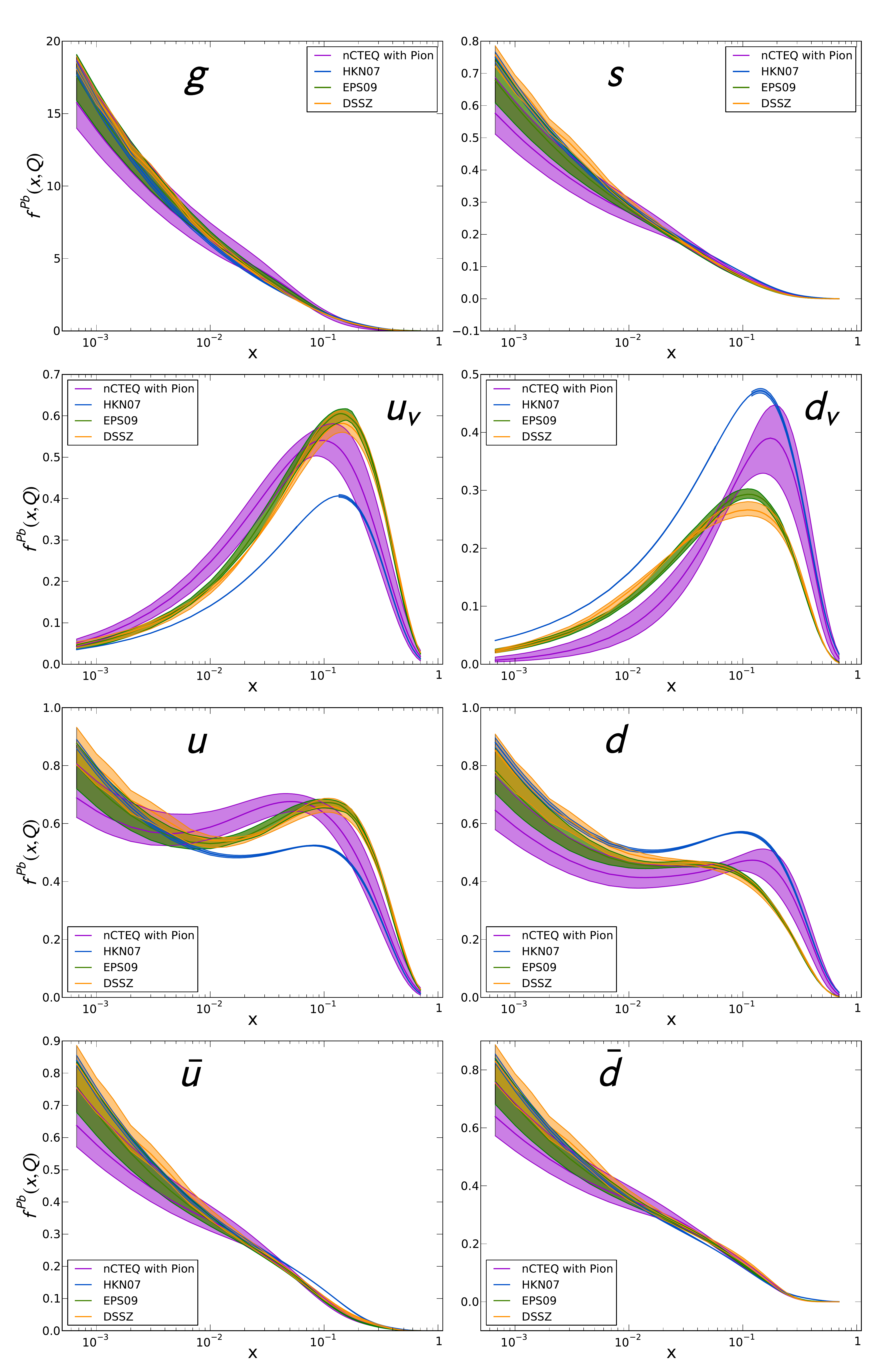}
\caption{Comparison of \ncteq ``pion fit'' (violet)
with results from other groups:
EPS09~\cite{Eskola:2009uj} (green),
DSSZ~\cite{deFlorian:2011fp} (orange),
HKN07~\cite{Hirai:2007sx} (blue).
The left panel shows nuclear modification factors for lead and the right panel the actual
PDFs of a proton bound in lead, both at scale of 10 GeV.}
\label{fig:compar_PDFs}
\end{figure*}

In figure~\ref{fig:compar_PDFs} we compare the \ncteq fit including pion data
with fits from other groups providing nuclear PDFs~\cite{Eskola:2009uj,deFlorian:2011fp,Hirai:2007sx}.
We can see a good agreement between the \ncteq and both EPS09~\cite{Eskola:2009uj}
and DSSZ~\cite{deFlorian:2011fp} PDFs in case of gluon and sea quarks.
However there is a big difference in case of the valence distributions, especially
for $d$-valence. This difference is well understood, and it is related to the
assumptions for the parametrization used during the fit. In both the EPS09 and
DSSZ approaches the nuclear correction for $u$-valence and $d$-valence PDFs is
universal which means that the two are tied together, whereas in the \ncteq
framework we treat them as independent (having separate fitting parameters).
If we look at the valence distributions obtained by the HKN~\cite{Hirai:2007sx}
group, where $u$-valence and $d$-valence are also treated as independent,
we can see that the \ncteq distributions lie in between those of HKN and EPS09 or DSSZ.
This behavior clearly illustrates the fact that there is not enough data to properly
constrain the nuclear PDFs, and shows that error PDFs obtained by each of the groups
do not fully quantify the actual uncertainties related with the nuclear PDFs.
Errors obtained in our analysis are bigger than in PDFs from other groups, which
can be seen as a more conservative representation of the uncertainties.
However, for a most realistic estimate of the actual PDF-related uncertainties we should combine errors
obtained by different groups.
%
%

\section{Comparison with data and some LHC predictions}
\label{sec:pred}
\begin{figure}
\begin{center}
\subfloat[]
{
\includegraphics[width=0.48\textwidth]{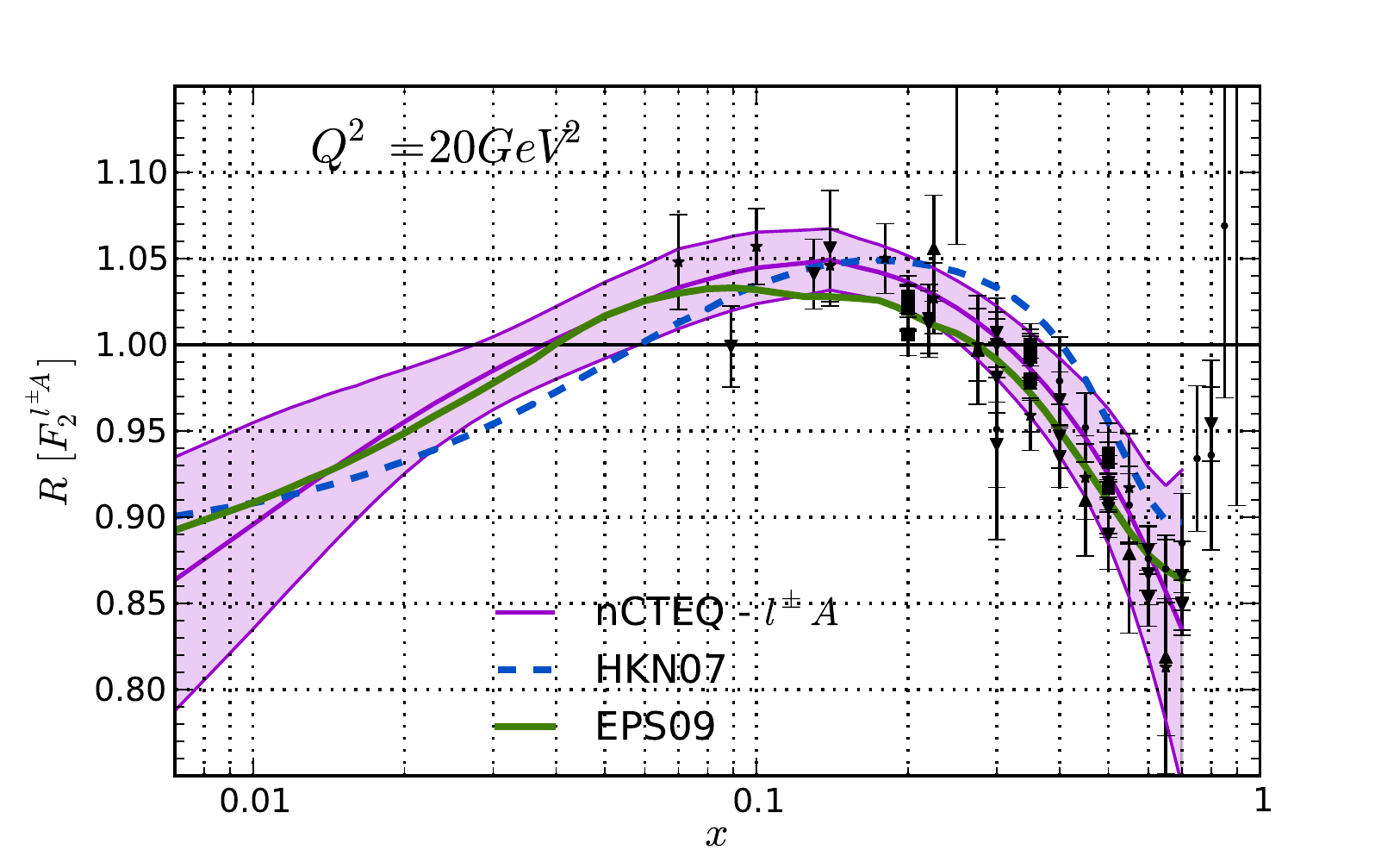}
\label{subfig:F2}
}
\quad
\subfloat[]
{
\includegraphics[width=0.45\textwidth]{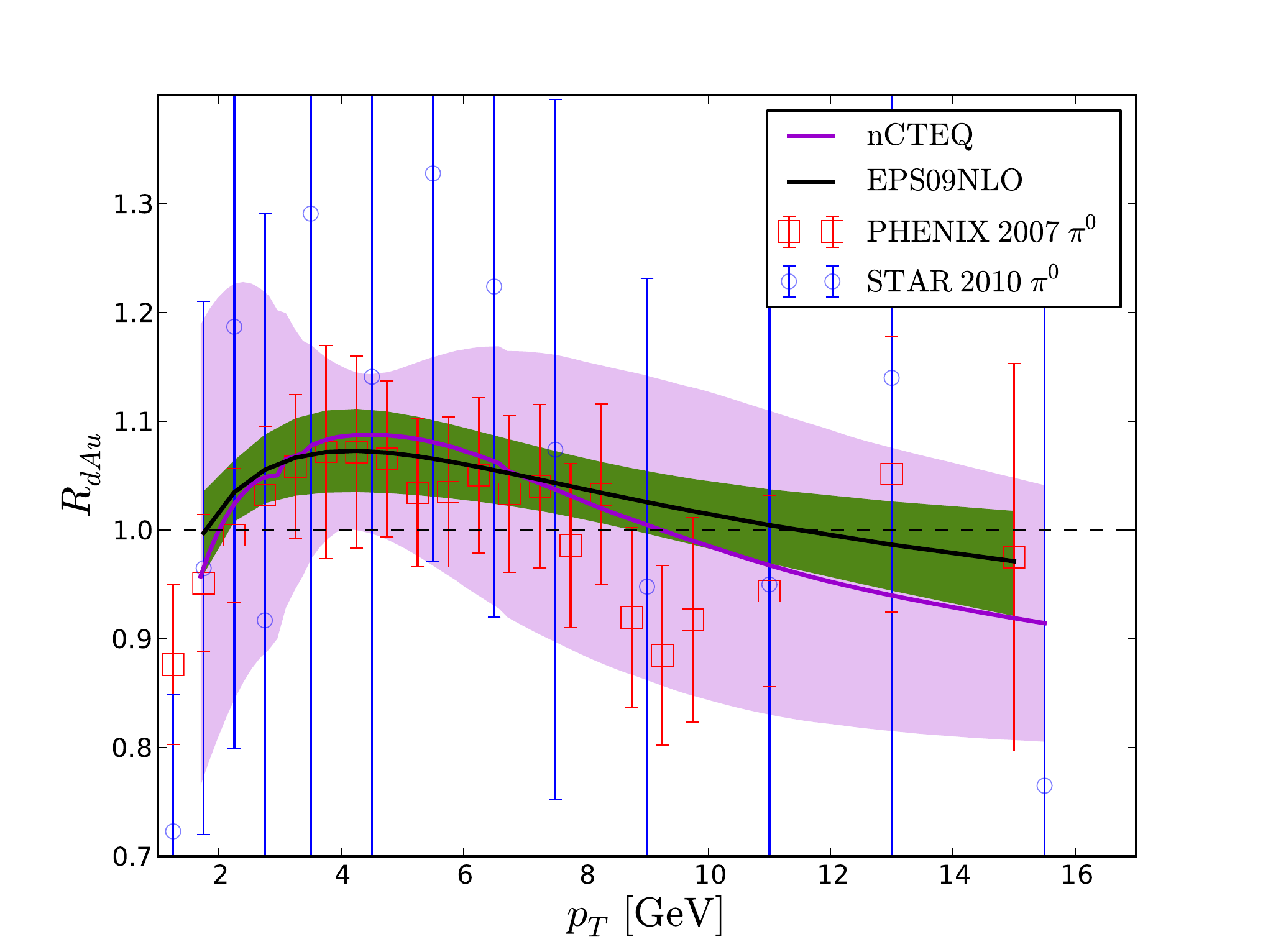}
\label{subfig:pion}
}
\caption{(a) Ratio of $F_2$ structure functions for iron and deuteron calculated
with \ncteq ``pion fit'' overlaid with fitted data and results from EPS09 and HKN07;
(b) $R_{\text{dAu}}^{\pi}$ computed with \ncteq and EPS09~\cite{Eskola:2009uj}
PDFs overlaid with PHENIX~\cite{Adler:2006wg} and STAR~\cite{Abelev:2009hx} data.}
\label{fig:data}
\end{center}
\end{figure}

Here we show how well our fit describes the fitted data.
As an example, in figure~\ref{subfig:F2}, we show data for the ratio of $F_2$
structure functions~\cite{Bodek:1983qn,Gomez:1993ri,Dasu:1993vk}
for iron and deuteron overlaid with our predictions (violet band)
and with central predictions of EPS09 and HKN07 fits.
We can see that our PDFs provide a very good description of the data,
also the error bands quite precisely reflect the errors of the data.

In figure~\ref{subfig:pion} we look at the pion data that was used during our analysis.
In this figure we show the nuclear modification of the pion yield:
\begin{equation}
R_{\text{dAu}}^{\pi} = \frac{\tfrac{1}{2A}d^2\sigma_{\pi}^{\text{dAu}}/dp_T dy}{d^2\sigma_{\pi}^{\text{pp}}/dp_T dy}
\end{equation}
measured in RHIC and compare it with predictions computed using our fit and
using the EPS09 fit (which also used some of these data).
We can see that the experimental errors are very large, especially in case of the
STAR data, and this is reflected in our error band (violet). Comparing with EPS09
we can see that our central values are close but EPS09 features much smaller
uncertainties. This is caused by the fact that in EPS09 analysis only the PHENIX data was
used (with much lower experimental errors) and because the PHENIX data was included with
weight of 20, magnifying its impact, whereas we used a weight of 1.

\begin{figure*}[t]
\centering{}
\includegraphics[width=0.42\textwidth]{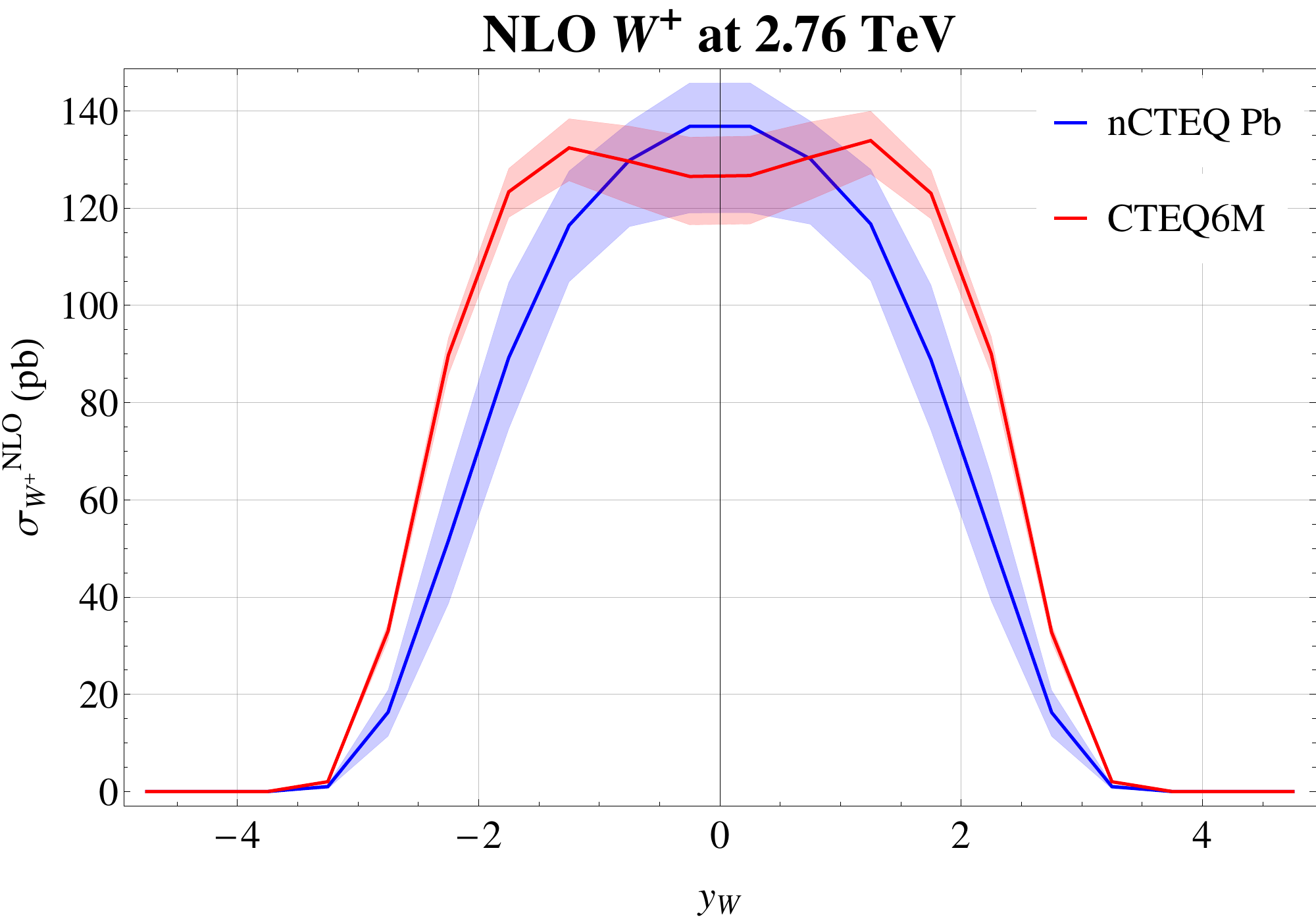}
\quad\quad
\includegraphics[width=0.42\textwidth]{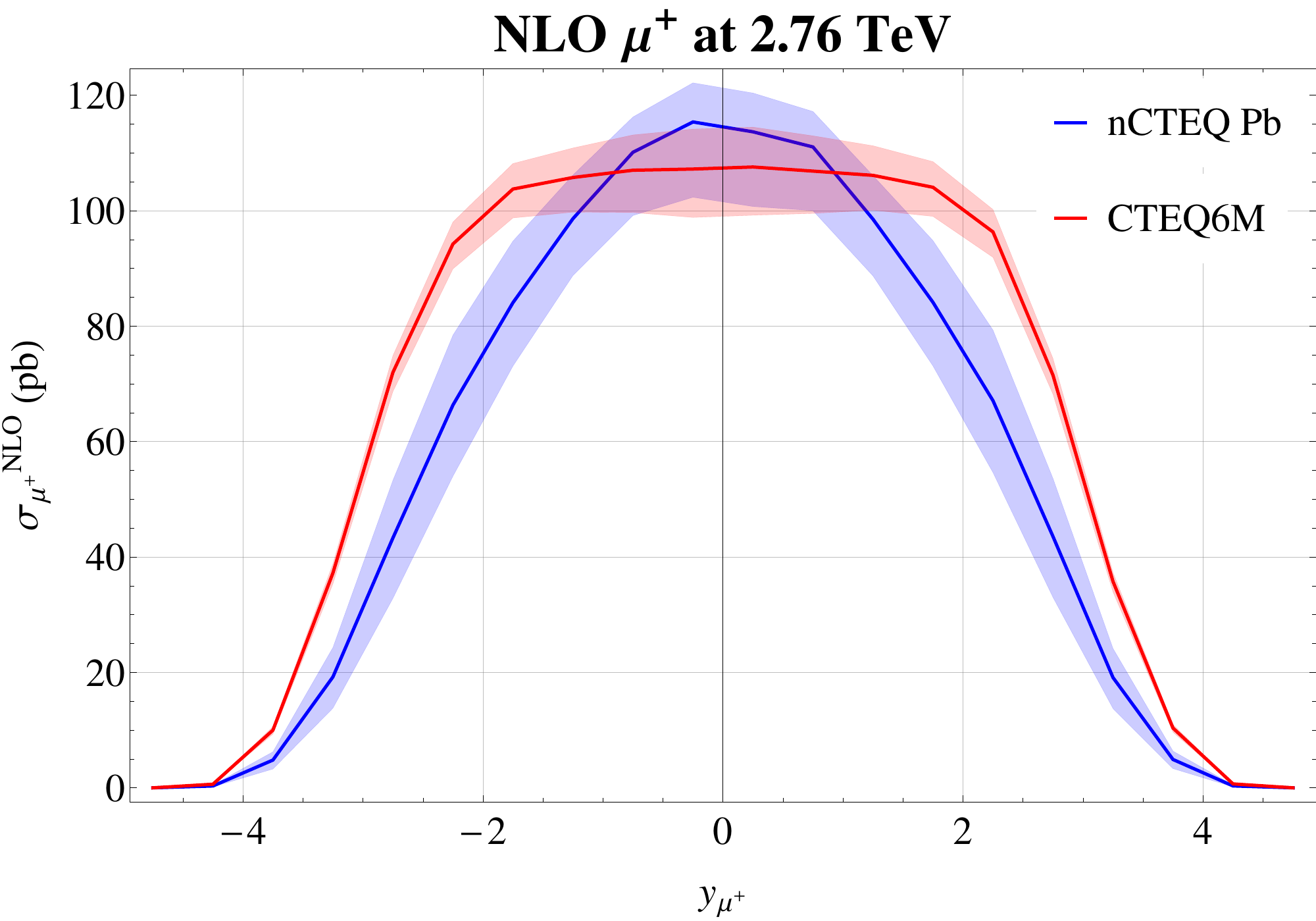}
\caption{Comparison of NLO $W^+$ (left) and $\mu^+$ (right) rapidity
distributions in case of proton (red) and lead (blue) collisions with
$\sqrt{s}=2.76$ TeV.}
\label{fig:Wp}
\end{figure*}

As a last thing we present predictions for $W^+$ production in lead collisions at
the LHC calculated at NLO using our new fit.%
    \footnote{For calculating NLO distributions for $W^+$ and $\mu^+$ the FEWZ
    code~\cite{Gavin:2010az} was used.}
In a recent ATLAS analysis~\cite{ATLAS-CONF-2013-106} the rapidity distributions
of the muon originating from the decay of $W^+$ boson was measured.
In this analysis free-proton PDFs were used to estimate a theoretical prediction
for this distribution, which was advocated to be sufficient due to the
substantial experimental errors and limited knowledge of the nuclear PDFs.
However, as can be seen in figure~\ref{fig:Wp} distributions obtained with the
free-proton PDFs (red) and with the nuclear PDFs (blue) are very different
(even for $\mu^+$), and we should become sensitive to these differences with more data.

%
%

\section{Conclusion}
\label{sec:conclusions}
We have presented a preliminary analysis of \ncteq nuclear PDFs introducing an error analysis
to our framework, and including pion production data that we have not
used before.

We performed a comparison with the results of other groups~\cite{Eskola:2009uj,deFlorian:2011fp,Hirai:2007sx}
and found our PDFs to be in reasonably good agreement with them. Also the 
discrepancies occurring in the valence distributions are well understood.
The estimates of errors provided by our fit  are larger than those of other
groups. The possible reason for this is related to the number of free
parameters used in the fit and with the flexibility of the parametrization
itself.

The presented results are still preliminary. The official \ncteq release
(providing PDF grids) will appear later this year.

\section*{Acknowledgments}
This work was partially supported by the U.S. Department of Energy
under grant {DE-FG02-13ER41996}, and the Lighter Sams
Foundation.

\bibliographystyle{utphys_spires}
\bibliography{biblio}


\end{document}